\documentclass[pdflatex,sn-mathphys-num]{sn-jnl}

\usepackage{graphicx}
\usepackage{multirow}
\usepackage{amsmath,amssymb,amsfonts}
\usepackage{mathrsfs}
\usepackage{xcolor}
\usepackage{textcomp}
\usepackage{booktabs}
\usepackage{algorithm}
\usepackage{algorithmicx}
\usepackage{algpseudocode}
\usepackage{listings}
\usepackage{array}

\begin{document}

%%%%%%%%%%%%%%%%%%%%%%%%%%%%%%%%%%%%%%%%%%%%%%%%%
% Title & Authors
%%%%%%%%%%%%%%%%%%%%%%%%%%%%%%%%%%%%%%%%%%%%%%%%%

\title[Toward Artificial Intelligence–Enabled Earth System Coupling]
{Toward Artificial Intelligence–Enabled Earth System Coupling}

\author*[1]{\fnm{Maria} \sur{Kaselimi}}\email{mkaselimi@noa.gr}

\author[1]{\fnm{Anna} \sur{Belehaki}}

\affil*[1]{\orgname{Institute for Astronomy, Astrophysics, Space Applications and Remote Sensing, National Observatory of Athens
}, \orgaddress{\city{15236 Penteli}, \country{Greece}}}

\keywords{Earth system, interactions, coupling, physics-informed machine learning, graph neural networks, causal discovery}

%%%%%%%%%%%%%%%%%%%%%%%%%%%%%%%%%%%%%%%%%%%%%%%%%
% Abstract
%%%%%%%%%%%%%%%%%%%%%%%%%%%%%%%%%%%%%%%%%%%%%%%%%

\abstract{
Coupling constitutes a foundational mechanism in the Earth system, regulating the interconnected physical, chemical, and biological processes that link its spheres. This review examines how emerging artificial intelligence (AI) methods create new opportunities to enhance Earth system coupling and address long-standing limitations in multi-component models. Rather than surveying next-generation modelling efforts broadly, we focus specifically on how state-of-the-art AI techniques can strengthen cross-domain interactions, support more coherent multi-component representations, and enable progress toward unified Earth system frameworks. The scope extends beyond climate models to include any modelling system in which Earth spheres interact. We outline emerging opportunities, persistent limitations, and conceptual pathways through which AI may enhance physical consistency, interpretability, and integration across domains. In doing so, this review provides a structured foundation for understanding the role of AI in advancing coupled Earth system modelling.
}

\maketitle

%%%%%%%%%%%%%%%%%%%%%%%%%%%%%%%%%%%%%%%%%%%%%%%%%
% 1. Introduction
%%%%%%%%%%%%%%%%%%%%%%%%%%%%%%%%%%%%%%%%%%%%%%%%%

\section{Introduction}

\label{sec:introduction}
The Earth system is commonly described as comprising four major, dynamically interacting components—the atmosphere, hydrosphere, geosphere, and biosphere. In many frameworks, these are complemented by additional subsystems, notably the cryosphere (snow, sea ice, glaciers, and permafrost) and the anthroposphere, which encompasses human activities and socio-environmental processes. These spheres do not operate in isolation; rather, they are tightly coupled through the continuous exchange of energy, mass, momentum, and chemical species. Such exchanges drive variability and feedbacks across a wide range of spatial and temporal scales, shaping phenomena from weather and hydrology to climate dynamics and biogeochemical cycles. Within this context, coupling refers to the explicit representation of these bidirectional or multilateral interactions in numerical models and data-driven frameworks. Coupling captures the mechanisms through which perturbations in one subsystem propagate to others, altering their state and behavior and ultimately modulating the evolution of the Earth system as a whole \cite{Eyring2021_ESM_Overview, Pan2025_ESR}.

Traditional earth modelling rely on a modular coupling architecture in which separately developed component models—such as the atmosphere, ocean, land surface, and cryosphere—are interconnected through a central coupler \cite{Pan2025_ESR}. This design has enabled major advances in climate and weather prediction, but several persistent limitations remain. High-resolution coupled simulations are computationally expensive, restricting ensemble size and limiting systematic exploration of uncertainty. Structural and parametric errors originating in one component can propagate across interfaces, producing long-term drifts in energy, water, or carbon budgets \cite{Best2011_JULES}. Scale mismatches between resolved dynamics and parameterized subgrid processes complicate the representation of interfacial fluxes \cite{Hourdin2017_ParamTuning}, while heterogeneous spatial grids, remapping algorithms, and asynchronous time stepping introduce additional numerical inconsistencies \cite{zhang2020stability}. The joint assimilation of multi-sphere observations—necessary to maintain physically consistent coupled states—further increases methodological complexity and computational cost \cite{Penny2017_CoupledDA, Carrassi2018_CoupledDA}. These limitations underscore the need for complementary approaches capable of representing coupling processes more flexibly while preserving physical integrity.

Recent advances in artificial intelligence (AI) and machine learning (ML) have opened new avenues for improving Earth system prediction, particularly through the emergence of large-scale foundation models. Early Earth-observation and atmospheric foundation models demonstrated impressive capabilities in learning coherent, multi-variable representations from vast and heterogeneous datasets. Examples include NASA–IBM’s Prithvi-EO-2.0 \cite{Prithvi2024}, Microsoft’s TerraFM \cite{TerraFM2025}, and Google DeepMind’s TerraMind \cite{TerraMind2025}, which provide powerful multimodal encodings of satellite observations, as well as atmospheric models such as ClimaX \cite{Nguyen2023_ClimaX}, which learns variable-agnostic representations of reanalysis fields and supports a wide range of downstream tasks. These systems demonstrated that foundation models can extract transferable structure from large geoscientific datasets and enable flexible generalization across variables, sensors, and regions. 

Building on these advances, recent efforts have begun to extend foundation models beyond single domains toward architectures that can directly represent multi-sphere interactions. Coupling-oriented foundation models—including Microsoft’s Aurora \cite{Bodnar2025_Aurora}, European Centre for Medium-Range Weather Forecasts (ECMWF) AI Forecasting System (AIFS) \cite{lang2024aifs}, and ORBIT \cite{Wang2024_ORBIT}—are trained on integrated atmospheric, land, ocean-wave, and climate-mode datasets, enabling them to capture cross-component dependencies and teleconnection patterns within a unified framework. Their distinguishing feature is the ability to approximate aspects of coupling. Importantly, the value of these models arises not from AI alone but from the combination of large-scale learning with embedded domain knowledge, including physical constraints, curated multi-sphere training corpora, and architectures designed to respect geophysical structure \cite{bordoni2025futures}. This integration makes coupling-oriented foundation models a promising complementary pathway for representing the physically mediated exchanges that govern the behavior of the atmosphere, ocean, land, cryosphere, and human systems.

At the same time, a growing body of research is moving beyond purely data-driven strategies toward AI approaches that explicitly integrate domain knowledge, physical structure, and meaningful interactions among system components. Physics-informed ML and hybrid modelling frameworks embed conservation laws and process-based constraints, ensuring that learned representations remain consistent with established scientific principles. Neural operators learn continuous dynamical mappings and can be combined into multi-component surrogates that respect underlying functional relationships. Graph neural networks (GNNs) provide a natural way to encode spatial, temporal, and cross-variable interactions on irregular meshes and heterogeneous domains, while causal discovery and explainable AI offer complementary tools for identifying directional influences and revealing the structure of complex system behaviour. These emerging directions reflect an important shift in AI research—from treating Earth system prediction as a pattern-recognition problem to designing architectures that capture mechanisms, dependencies, and scientifically grounded reasoning~\cite{tuia2021toward}.

The purpose of this review is to synthesize and organize these emerging developments with explicit focus on coupling. Rather than surveying AI in Earth sciences broadly, we concentrate on methods whose primary role is to represent, emulate, constrain, or interpret interactions across components of the Earth system. Building on the physical perspective of coupling introduced in Section~\ref{sec:background} and the description of traditional coupling frameworks in Section~\ref{sec:traditional_coupling}, then in Section~\ref{sec:ai_coupling} we examine how AI can be integrated into these infrastructures and what new risks and opportunities this integration entails.

\begin{figure*}[h!]
    \centering
\includegraphics[width=0.8\textwidth]{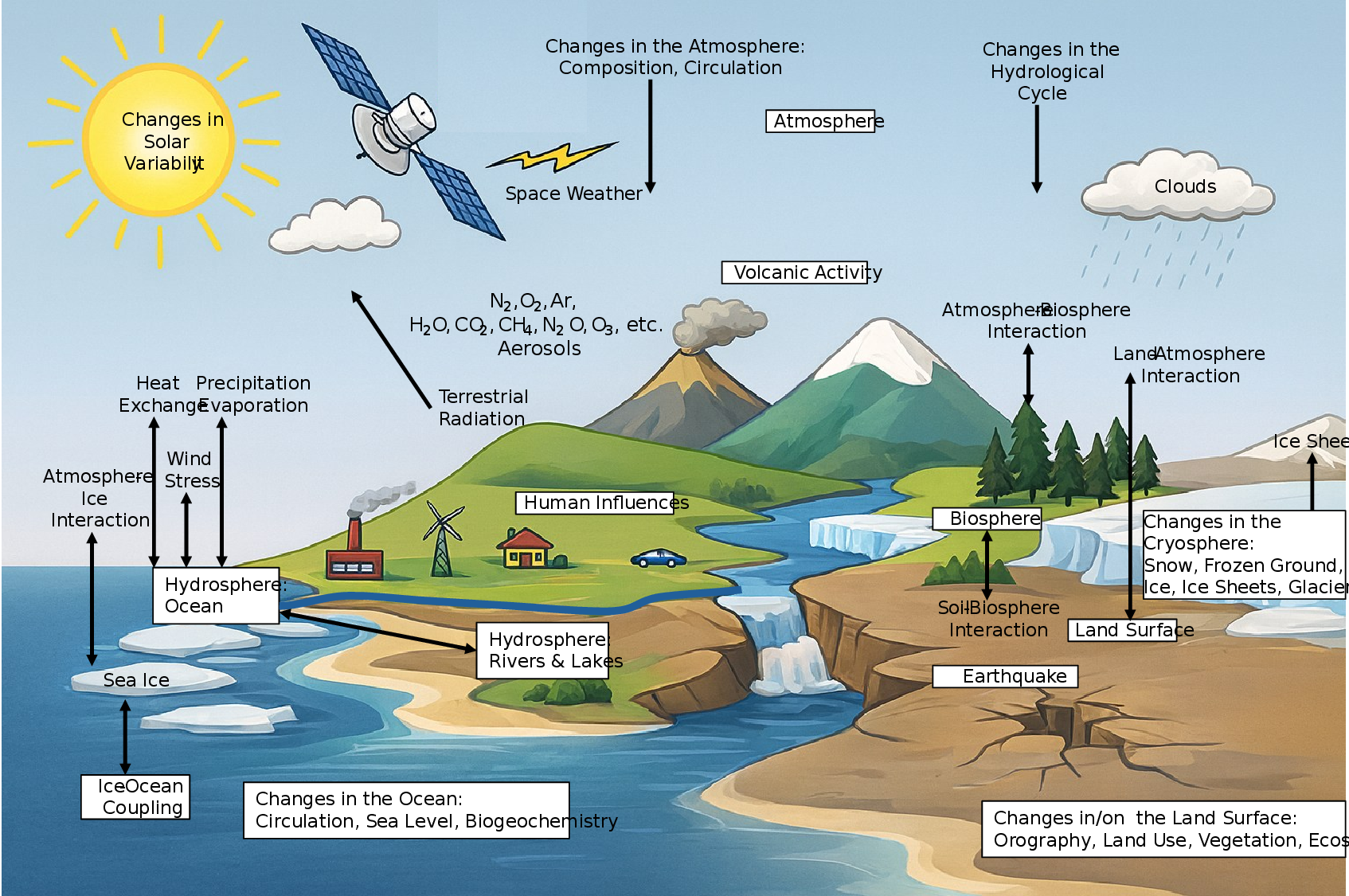}
    \caption{Earth System Interaction Diagram}
    \label{fig:climate-system-interaction}
\end{figure*}

Specifically, the objectives of this review are to:
\begin{itemize}
    \item Characterize the role of coupling in Earth system modelling and delineate the numerical and conceptual limitations of traditional coupling frameworks.
    \item Provide a detailed exploration on AI approaches for Earth system coupling, including interface and flux emulators, neural-operator-based surrogates, graph-based interaction models, multi-domain foundation models, and causal/interpretability methods.
    \item Assess the benefits and limitations of these AI approaches in coupling contexts, with emphasis on physical consistency, long-term stability, predictive skill, computational efficiency, and uncertainty quantification.
    \item Highlight the role of multi-component observational datasets---including satellite, in-situ, reanalysis, and socioeconomic data---in enabling AI-based coupling solutions and multi-domain data fusion.
    \item Outline evaluation strategies and open research directions for integrating AI-based couplers into next-generation ESMs and digital twin frameworks.
\end{itemize}

%%%%%%%%%%%%%%%%%%%%%%%%%%%%%%%%%%%%%%%%%%%%%%%%%
% 2. Scope and Boundaries
%%%%%%%%%%%%%%%%%%%%%%%%%%%%%%%%%%%%%%%%%%%%%%%%%

\section{Scope and Boundaries of This Review}
This review does not aim to provide a comprehensive assessment of next-generation Earth System Models (ESM) in their entirety, nor does it attempt to catalogue the full landscape of advances in climate modelling. Instead, its scope is intentionally focused on coupling—that is, on the physical and computational mechanisms that mediate interactions across the atmosphere, hydrosphere, geosphere, biosphere, and cryosphere. By design, this review concentrates on how emerging AI approaches can strengthen these interactions, address long-standing numerical limitations, and offer conceptual pathways toward more unified and physically coherent multi-component models. Because our goal is to examine coupling across all Earth system domains, the discussion goes beyond climate-centric ESMs to include coupled models in biogeochemistry, human–Earth systems, near-Earth space, and lithosphere–atmosphere–ionosphere interactions, as illustrated in the broader system context (see Fig. 1). The  focus of this review has been deliberately restricted to AI methods whose primary function is to emulate, enhance, or interpret coupling processes across Earth system components. As a result, several related areas—such as AI for pure Earth observation, downscaling, extremes prediction, or parameterization design—are discussed only when they directly inform coupling mechanisms. Given the breadth of the topic and the rapid pace of methodological innovation, this review should therefore be viewed as a conceptual synthesis oriented around coupling-specific challenges and opportunities, rather than as an exhaustive survey of next-generation Earth system modelling.

%%%%%%%%%%%%%%%%%%%%%%%%%%%%%%%%%%%%%%%%%%%%%%%%%
% 3. Background: Earth System Coupling
%%%%%%%%%%%%%%%%%%%%%%%%%%%%%%%%%%%%%%%%%%%%%%%%%

\section{Background: Earth System Coupling}
\label{sec:background}

\subsection{What is Coupling?}

The Earth system comprises interacting components---atmosphere, ocean, land surface (including vegetation and soils), cryosphere, and biogeochemical cycles---that exchange energy, mass, and momentum across their interfaces. In this review, we use \emph{coupling} to refer to the representation of these bidirectional or multilateral interactions in models and data-driven frameworks, with an emphasis on how perturbations in one component propagate through the others and modify their state \cite{Eyring2021_ESM_Overview, Valcke2012_CouplingESM}. 

At large scales, coupling emerges from fundamental conservation principles. Radiative imbalances generate atmospheric and oceanic circulations. Hydrological and biogeochemical cycles transport water and carbon across the system. Mechanical stresses link the solid Earth with surface and atmospheric processes. Well-known examples include the El~Ni\~no--Southern Oscillation (ENSO), the North Atlantic Oscillation (NAO), and soil-moisture–precipitation feedbacks, all of which illustrate how anomalies in one region or subsystem can induce remote atmospheric and hydrological responses \cite{Trenberth2020_Teleconnections,  Dirmeyer2021_LandAtm_Coupling}. 

Although these classical domains form the backbone of traditional Earth System Models, recent research shows that coupling extends beyond the atmosphere–ocean–land–cryosphere system. Interactions among the lithosphere, atmosphere, and ionosphere, such as radon-induced ionization, thermal anomalies, and perturbations in ionospheric electron content, exemplify an emerging vertical coupling pathway \cite{Pulinets2018_LAIC_Review}. Ocean–biogeochemistry interactions, which link circulation with nutrient supply, ecosystem productivity, and carbon uptake, represent another domain in which physical and biological processes are tightly coupled \cite{doney2012climate}. Human activity also forms a two-way coupling with the physical climate through land-use change, emissions, irrigation, and energy use, all of which modify surface fluxes and atmospheric composition while simultaneously responding to environmental change \cite{Calvin2019_CHESM, Douville2021_HumanClimate}. Coupling processes also extend into near-Earth space, where solar forcing links the magnetosphere, ionosphere, and thermosphere and influences geomagnetic activity and satellite drag \cite{Camporeale2019_SpaceWeatherAI}. Recent space-to-Earth studies further demonstrate that coupling processes extend into the ionosphere–thermosphere–magnetosphere system. For example, comprehensive space-weather system overviews highlight how solar-wind–magnetosphere interactions modulate ionospheric electrodynamics, thermospheric heating, and satellite-drag variability, forming a tightly connected geospace–Earth coupling chain \cite{beedle2022user}.

In this broader perspective, coupling encompasses the full set of dynamical pathways—physical, chemical, biological, and anthropogenic—through which information, fluxes, and feedbacks are transmitted across components of the Earth system. For AI-enabled modeling, these couplings define the structural constraints and cross-domain dependencies that learning-based systems must emulate, predict, or infer in order to represent the integrated, multiscale behaviour of the Earth system.

\subsection{Coupling Interfaces and Fluxes}

Traditional Earth System Models and emerging AI-based approaches both rely on a set of key coupling interfaces:

\textbf{Atmosphere--ocean:} At the air--sea interface, the atmosphere provides wind stress, precipitation, radiative fluxes, and gas transfer conditions, while the ocean returns sea-surface temperature (SST), currents, and surface heat and freshwater fluxes. These exchanges regulate the global energy and water budgets, control variability modes such as ENSO, and shape the location of storm tracks and monsoon systems \cite{Levine2021_SeaIceLoss}. 

\textbf{Land--atmosphere:} The land surface and terrestrial biosphere exchange latent and sensible heat, CO$_2$, water vapour, and momentum with the atmosphere. Soil moisture, vegetation structure, snow cover, and land-use practices modulate surface albedo and roughness, influencing boundary-layer development, convection, and precipitation \cite{Bonan2019_LandSurfaceReview, Lawrence2022_ESM_Land}. Land--atmosphere coupling plays a central role in amplifying heatwaves, droughts, and compound extremes \cite{Dirmeyer2021_LandAtm_Coupling}. 

\textbf{Ocean--ice and cryosphere coupling:} Sea ice modulates atmosphere--ocean interactions through its albedo, insulating properties, and freshwater fluxes. The growth and melt of sea ice regulate ocean stratification and circulation, while snow and ice cover influence radiative budgets and boundary-layer stability. These processes are critical for polar climate feedbacks and long-term sea-level projections \cite{Hunke2015_CICE, Vancoppenolle2012_LIM}. 

These classical interfaces demonstrate how physical fluxes and feedbacks bind the components of the Earth system. Increasingly, however, additional domains of coupling are recognized, expanding the scope of data-driven and AI-based approaches.

\subsection{Emerging Cross-Domain Coupling Beyond Traditional ESMs}
\label{sec:emerging_coupling}

Beyond the atmosphere–ocean–land–cryosphere system, several frontier domains exhibit coupling processes that extend the traditional boundaries of ESM-based frameworks. These domains operate across disparate spatial and temporal scales and rely heavily on heterogeneous observations, making them natural targets for AI-based modeling, anomaly detection, and multi-modal data fusion.

\textbf{Lithosphere–Atmosphere–Ionosphere Coupling (LAIC):}
LAIC describes how lithospheric processes---including crustal stress accumulation, rock fracturing, and radon or gas emissions---can generate perturbations in atmospheric conductivity, thermal structure, and ionospheric electron content. Observables include ionospheric total electron content (TEC) anomalies detected by Global Navigation Satellite Systems (GNSS), air-ionization changes induced by radon release, and atmospheric infrared signatures \cite{Pulinets2018_LAIC_Review, Parrot2024_LAIC_AI}. Because these signals are weak and embedded in noise, AI methods have been used to detect anomaly patterns, classify precursor signals, and fuse multiple sparse datasets across the lithosphere, atmosphere, and ionosphere. Modern LAIC studies increasingly use AI to extract weak precursory signals embedded in ionospheric and atmospheric variability. Deep learning frameworks have successfully identified TEC perturbations preceding large earthquakes, providing new evidence for upward propagating coupling mechanisms and highlighting the utility of ML in isolating LAIC signatures from noise \cite{tsai2022deep}. These approaches reinforce the idea that coupling processes extend from the solid Earth through the atmosphere into ionospheric plasma, forming a vertically integrated system observable with modern geospace monitoring infrastructure \cite{belehaki2025integrating, laurenza2025course}.

\begin{sidewaystable}[t]
\caption{Traditional numerical coupling challenges and illustrative AI-enabled opportunities. }
\label{tab:coupling_ai_challenges}
\centering
\small
\setlength{\tabcolsep}{4pt}
\renewcommand{\arraystretch}{0.95}
\begin{tabular}{p{0.10\textwidth} p{0.34\textwidth} p{0.45\textwidth}}
\hline
\textbf{Challenge} &
\textbf{Representative issues in numerical ESMs} &
\textbf{Illustrative AI-enabled opportunities} \\
\hline
\textbf{Physical consistency across interfaces} &
Component-specific parameterizations, reference conventions, and bulk algorithms lead to inconsistencies in surface fluxes and imperfect conservation of energy, mass, and moisture across domains \cite{gross2018physics,Warner2008_MCT}. &
Physics-informed interface and flux emulators that enforce conservation constraints; learned conservative remapping operators; AI-assisted tuning of interface schemes to minimize budget residuals and flux mismatches \cite{yuval2020stable,Beucler2021_PhysicsML}. \\

\textbf{Scale mismatch and resolution gaps} &
Atmosphere, ocean, land, and cryosphere operate on different spatial grids and time steps; key subgrid processes (convection, turbulence, eddies, heterogeneity) are parameterized differently across components, distorting multiscale coupling \cite{Hourdin2017_ParamTuning}. &
Multi-resolution neural operators and super-resolution architectures for mapping between heterogeneous grids; foundation models that learn shared, scale-aware latent spaces across variables and domains; and surrogate parameterizations trained jointly on multi-sphere datasets to represent cross-scale interactions consistently \cite{Bodnar2025_Aurora}. \\

\textbf{Numerical stability \& feedback issues} &
Asynchronous coupling, temporal interpolation \& non-conservative remapping induce spurious oscillations or unstable feedbacks, degrading long-term stability of coupled integrations \cite{lemarie2015analysis}. &
Stability-aware surrogate couplers trained with long-horizon rollouts and physical regularization; AI-guided design of coupling intervals and interface formulations; emulation of fast processes to reduce stiffness in strongly coupled regimes \cite{camps2025artificial}. \\

\textbf{Incompatible physical parameterizations } &
Different treatments of boundary-layer turbulence, cloud--radiation interactions, land--surface processes, and sea-ice thermodynamics create incoherent feedbacks and distorted interface budgets. &
Hybrid schemes in which AI learns residual corrections between component parameterizations; shared cross-domain parameterizations derived from multi-sphere training (e.g., graph-based or foundation models); emulator layers that harmonize fluxes across distinct physics packages \cite{Bonan2019_LandSurfaceReview}. \\

\textbf{Sparse or biased observations at interfaces} &
Limited in situ and satellite constraints for air--sea exchanges, mixed-layer dynamics, soil moisture, and cryospheric transitions hinder calibration and validation of interface processes. &
Physics-informed fusion methods that embed conservation laws in multimodal integration of reanalyses, satellite, and in situ data \cite{raissi2019physicsinformed}; coupled data assimilation with ML-aided bias corrections and adaptive error covariances; physics-guided residual emulators for gap filling in sparse observations. \\

\textbf{Computational cost and synchronization overhead} &
High-resolution coupling requires frequent exchange of fields, conservative remapping, and diagnostics, which increase communication cost and limit ensemble size and scenario exploration. &
Fast surrogate models for expensive interface physics and subgrid schemes; neural-operator surrogates for slow components (e.g., eddy-rich ocean, sea ice) to enable higher coupling frequency; AI-driven emulation of reduced-order coupled dynamics for large ensembles \cite{supto2025next}. \\

\textbf{Bias propagation across components} &
Systematic errors in one subsystem (e.g., SST, soil moisture, albedo) propagate through flux exchanges, affecting circulation, hydrology, and cryosphere, and complicating attribution of coupled biases. &
Constrained bias-correction networks that operate on coupled fluxes and tendencies; causal-discovery and attribution tools to diagnose pathways of bias propagation; residual emulators trained to reduce long-term drift while preserving variability and teleconnections \cite{feng2025disentangling}. \\

\textbf{Integration of new components and domains} &
Adding biogeochemistry, human--Earth interactions, and near-Earth space processes increases dimensionality and heterogeneity, complicating interface design and risking new inconsistencies. &
Modular AI surrogates for emerging components that expose standardized coupling variables; multi-domain foundation models that learn shared latent spaces across variables; graph-based architectures that flexibly represent new interaction pathways; Causal discovery \cite{tuia2021toward}. \\
\hline
\end{tabular}
\end{sidewaystable}

\textbf{Human--Earth coupling:}
Human activities---including land-use change, irrigation, urbanization, emissions, and resource extraction---alter surface fluxes, hydrological pathways, and boundary-layer properties. Coupled Human–Earth System Models (CHESMs) treat socioeconomic dynamics, energy systems, and land-use decisions as interacting components of the Earth system \cite{Calvin2019_CHESM, Douville2021_HumanClimate}. AI contributes by learning human-driven flux patterns from remote sensing, integrating socioeconomic and environmental datasets, detecting emergent feedbacks, and emulating human-system components in CHESMs.

\textbf{Ocean–biogeochemistry and ecosystem coupling:}
Marine ecosystems interact strongly with physical circulation through nutrient transport, plankton growth, carbon uptake, and mixed-layer dynamics. These nonlinear biogeochemical feedbacks remain challenging due to sparse observations and high model complexity \cite{doney2012climate}. AI techniques—including graph neural networks, neural operators, and multimodal fusion models—assist in emulating biogeochemical fluxes, integrating satellite-derived chlorophyll and nutrient data, and discovering ecosystem–circulation relationships.

\textbf{Magnetosphere–ionosphere–thermosphere coupling:}
Solar wind forcing induces a cascade of interactions across the magnetosphere, ionosphere, and thermosphere, influencing ionospheric conductivity, geomagnetic activity, and satellite drag. Hybrid physical–AI approaches are now used to predict geomagnetic storms, infer magnetosphere–ionosphere coupling pathways, and fuse space-weather observations \cite{Camporeale2019_SpaceWeatherAI}. Recent AI-driven space-weather forecasting frameworks further demonstrate the strong interdependence between magnetospheric forcing, ionospheric electrodynamics, and thermospheric neutral responses. Evaluations of ML-based geospace forecasting tools emphasize their role in predicting geomagnetic disturbances, satellite-drag conditions, and radiation-belt variability, all governed by multi-stage coupling processes across Earth’s near-space environment \cite{thaker2025evaluating}. In parallel, system-level reviews of magnetically connected space-weather pathways synthesize how plasma transport, field-aligned currents, and Joule heating transmit perturbations across domains, providing a conceptual foundation for AI models targeting multi-region interactions \cite{beedle2022user}. AI methods have also been applied to disturbances arising from magnetosphere–ionosphere–thermosphere coupling: for example, neural-network-based prediction of Large-Scale Travelling Ionospheric Disturbances (LSTIDs)—a key manifestation of geomagnetic energy input into the upper atmosphere—shows that data-driven architectures can recover coupling pathways associated with auroral heating, thermospheric expansion, and ionospheric wave propagation. Themelis et al.\ (2025) employed a Temporal Fusion Transformer (TFT), a multivariate sequence-to-sequence deep-learning model, to capture both short-term dynamics and long-range coupling dependencies between geomagnetic drivers and ionospheric responses, highlighting AI’s expanding ability to model upper-atmospheric responses to space-weather forcing \cite{themelis2025neural}. These systems extend Earth system coupling into the near-Earth space environment.

% These emerging domains demonstrate that coupling extends far beyond climate-centric ESMs. Their heterogeneous data streams and weak or intermittent signals make them natural targets for AI methods that excel at anomaly detection, cross-domain fusion, and learning from incomplete or noisy observations.

%%%%%%%%%%%%%%%%%%%%%%%%%%%%%%%%%%%%%%%%%%%%%%%%%
% 4. Traditional Approaches
%%%%%%%%%%%%%%%%%%%%%%%%%%%%%%%%%%%%%%%%%%%%%%%%%

\section{Traditional Approaches to Coupling and their Limitations}
\label{sec:traditional_coupling}
\subsection{Coupled Model Architectures}

Traditional ESMs couple specialized component models---atmospheric general circulation models (AGCMs), ocean general circulation models (OGCMs), land-surface and terrestrial biosphere models, and cryospheric models---through a central \emph{coupler} \cite{Eyring2021_ESM_Overview}. Each component solves its governing partial differential equations with its own numerics, resolution, and time step, while the coupler mediates the exchange of fluxes and state variables at specified coupling intervals.

Modern coupling infrastructures such as OASIS3--MCT, the Earth System Modeling Framework (ESMF), and the CPL7 coupler in CESM provide standardized mechanisms for conservative grid remapping, temporal synchronization, and diagnostics \cite{hill2004architecture, Craig2012_CPL7}. This modular architecture enables independent development but also makes it difficult to maintain strict conservation and consistency across domains.

Although these architectures were originally designed for climate-centric components, analogous traditional frameworks exist in several emerging coupling domains. For example, ocean–biogeochemistry interactions are commonly represented using nutrient–phytoplankton–zooplankton–detritus (NPZD) models or more complex biogeochemical tracers integrated within OGCMs \cite{doney2012climate}. Human–Earth interactions are typically incorporated through Integrated Assessment Models (IAMs) or economic–land–climate modules that exchange emissions, land-use changes, or resource constraints with simplified physical models \cite{Calvin2019_CHESM}. In the near-Earth space environment, magnetosphere–ionosphere coupling is traditionally modeled using physics-based space-weather systems that propagate solar wind forcing through empirical or semi-empirical ionospheric response functions \cite{Camporeale2019_SpaceWeatherAI}. Vertical lithosphere–atmosphere–ionosphere coupling (LAIC) is often represented using empirical radon–conductivity relationships or statistical models of ionospheric perturbations \cite{Pulinets2018_LAIC_Review}.

These examples illustrate that traditional coupling formulations, whether in climate-oriented ESMs or in extended geophysical and human–Earth domains, generally rely on explicit physical operators, empirical parameterizations, or simplified exchange rules. As in the core ESM components, these traditional frameworks face challenges related to sparse observations, incomplete physical understanding, and the need to represent multiscale feedback processes. 

\subsection{Challenges in Numerical Coupling}

Despite substantial progress in model development, numerical coupling in Earth system models faces several persistent structural limitations, as shown in table \ref{tab:coupling_ai_challenges}. A central challenge is maintaining physical consistency across component interfaces. Independently developed atmosphere, ocean, land, and cryosphere schemes often employ differing parameterizations, discretizations, or reference conventions, which can result in mismatches in surface exchange formulations and ultimately lead to flux imbalances or inconsistencies in conserved quantities \cite{Sellers1997_BiosphereAtmosphereCoupling,Best2011_JULES,Fairall2003_COAREFluxes}.

A second difficulty arises from scale mismatch between components. Each subsystem evolves on distinct spatial and temporal scales, and many essential coupling processes—including convection, boundary-layer turbulence, eddies, and heterogeneous land–surface exchanges—are parameterized rather than resolved. When components operate on different grids and timestep structures, these multiscale interactions may be distorted, damped, or spuriously amplified \cite{Hourdin2017_ParamTuning,Stensrud2007_ParametrizationWeather}.

Third, coupling can introduce numerical stability issues, particularly when using asynchronous updating, temporal interpolation, or non-conservative remapping. These procedures may generate oscillations or noise that disrupt dynamical balance at the interface and can degrade long-term stability \cite{zhang2020stability,Valcke2012_CouplingESM,gross2018physics}. Closely related are challenges associated with incompatible representations of physical processes across domain boundaries. Differences in cloud–radiation treatments, boundary-layer schemes, sea-ice thermodynamics, or land–surface formulations can alter feedbacks and distort energy and water budgets at the coupled interface \cite{edwards2020representation,Hourdin2017_ParamTuning}.

Coupling is further constrained by limited and heterogeneous observations in key interface regions. Air–sea turbulent exchanges, ocean mixed-layer variability, soil-moisture–boundary-layer interactions, and cryospheric transitions are often sparsely observed, hindering calibration, tuning, and evaluation of coupled behavior \cite{Dirmeyer2021_LandAtm_Coupling,doney2012climate}. In addition, high-resolution coupling imposes significant computational demands, as interactions across components require additional communication, interpolation, diagnostics, and synchronization overhead, which restricts ensemble size and limits the exploration of uncertainty \cite{Valcke2012_CouplingESM,hill2004architecture}.

A further structural limitation is bias propagation across subsystems. Errors arising in one component—such as SST biases, soil-moisture errors, or misrepresented albedo—can spread through flux exchanges and influence circulation, hydrology, and cryospheric evolution, complicating attribution and model tuning \cite{Brovkin2013_CoupledESM_Biogeochem}. Finally, modern Earth system models increasingly incorporate new components and interaction pathways, including biogeochemistry, human–Earth couplings, and near-Earth space processes. While scientifically essential, this expansion increases the dimensionality and heterogeneity of the coupling problem and introduces additional opportunities for inconsistencies or unstable feedbacks \cite{Calvin2019_CHESM, Camporeale2019_SpaceWeatherAI}.

Collectively, these challenges illustrate the structural complexity of numerical coupling and motivate complementary avenues for improvement. Emerging AI-based methods offer new opportunities to support traditional modeling frameworks by enhancing interface representations, strengthening physical consistency, and enabling more flexible treatment of multiscale interactions across Earth system components.

This perspective motivates the remainder of the review, which surveys recent AI methods for Earth system coupling, evaluates their benefits and limitations, and outlines how they may be integrated into next-generation ESMs and digital twin frameworks.

\begin{sidewaystable}[t]
\centering
\footnotesize
\caption{Comparison of single-system and coupling-oriented foundation models in Earth system science. Single-system models focus on Earth observation or atmospheric prediction without explicit physical coupling, whereas coupling-oriented models learn cross-domain dependencies by training on integrated multi-domain Earth system datasets.}
\label{tab:foundation-models-comparison}
\begin{tabular}{p{2.6cm} p{3.2cm} p{3.5cm} p{6.5cm}}
\hline
\textbf{Model} & \textbf{Type} & \textbf{Domains} & \textbf{Notes on Capabilities / Coupling Relevance} \\
\hline
Prithvi-EO-2.0~\cite{Prithvi2024} & Single-system (EO) & EO / multisensor imagery 
& Multimodal satellite representation; no dynamical prediction; no cross-domain physical interactions. \\

TerraFM~\cite{TerraFM2025} & Single-system (EO) & EO (Sentinel-1/2) 
& Large-scale multi-resolution EO model; strong spatial representation learning; not designed for physical coupling. \\

TerraMind~\cite{TerraMind2025} & Single-system (EO) & EO (9 modalities)
& Generative multimodal model; integrates diverse sensing modalities; not a predictive Earth system model. \\

ClimaX~\cite{Nguyen2023_ClimaX} & Single-system (Atmosphere) & Weather / reanalysis 
& Variable-agnostic atmospheric FM; learns atmospheric states but does not model interactions with ocean or land. \\
Aurora~\cite{Bodnar2025_Aurora} & Coupling-oriented & Atmosphere, waves, land, air quality
& Trained on multi-domain data; shared latent representation produces cross-domain attention; learns coupling-like relationships. \\

AIFS~\cite{lang2024aifs} & Coupling-oriented & Atmosphere, land, ocean waves 
& Unified forecasting system; reproduces emergent interactions (e.g., wave damping by sea ice) despite missing explicit components. \\

ORBIT~\cite{Wang2024_ORBIT} & Coupling-oriented & Atmosphere–ocean–climate modes
& Ultra-large foundation model targeting predictability; captures long-range temporal dependencies and teleconnections. \\
\hline
\end{tabular}

\end{sidewaystable}

%%%%%%%%%%%%%%%%%%%%%%%%%%%%%%%%%%%%%%%%%%%%%%%%%
% Foundation Model Comparison Table
%%%%%%%%%%%%%%%%%%%%%%%%%%%%%%%%%%%%%%%%%%%%%%%%%

% \begin{table*}[t]
% \centering
% \caption{Comparison of single-system and coupling-oriented foundation models in Earth system science.}
% \label{tab:foundation-models-comparison}
% \begin{tabular}{p{2.6cm} p{3.2cm} p{3.5cm} p{6.5cm}}
% \toprule
% \textbf{Model} & \textbf{Type} & \textbf{Domains} & \textbf{Notes} \\
% \midrule
% (Full table content preserved)
% \\
% \bottomrule
% \end{tabular}
% \end{table*}

%%%%%%%%%%%%%%%%%%%%%%%%%%%%%%%%%%%%%%%%%%%%%%%%%
% 5. AI Approaches for Coupling
%%%%%%%%%%%%%%%%%%%%%%%%%%%%%%%%%%%%%%%%%%%%%%%%%

\section{AI Approaches for Earth System Coupling}
\label{sec:ai_coupling}

Artificial intelligence is increasingly used to represent, emulate, or discover the interactions that couple the Earth system. Unlike traditional models, where interface fluxes and feedbacks are explicitly encoded through parameterizations and couplers, AI models can learn cross-domain interactions directly from data, while optionally imposing physical constraints. This section provides a unified overview of how different classes of AI methods contribute to the representation of coupled Earth system dynamics, highlighting both their capabilities and their limitations. A structural taxonomy summarizing these method classes, their purposes, and representative applications is provided in Table~\ref{tab:coupling_taxonomy_final}. It also outlines emerging good practices for integrating AI models into established Earth system modeling workflows—such as enforcing physical constraints, ensuring stability in long-range simulations, and harmonizing multi-domain datasets—and provides examples of how these approaches are increasingly being adopted within the Earth system science community.

%%%%%%%%%%%%%%%%%%%%%%%%%%%%%%%%%%%%%%%%%%%%%%%%%
% Taxonomy Table
%%%%%%%%%%%%%%%%%%%%%%%%%%%%%%%%%%%%%%%%%%%%%%%%%

%%%%%%%%%%%%%%%%%%%%%%%%%%%%%%%%%%%%%%%%%%%%%%%%%
% Subsections 5.1 - 5.5
%%%%%%%%%%%%%%%%%%%%%%%%%%%%%%%%%%%%%%%%%%%%%%%%%

\begin{sidewaystable}[t]
\centering
\caption{Structural AI Taxonomy for Representing and Enhancing Earth System Coupling. The taxonomy organizes AI approaches by their purpose, methodological subclasses, and representative coupled-domain applications.}
\renewcommand{\arraystretch}{1.1}
\begin{tabular}{p{2.8cm} p{3.5cm} p{4.8cm} p{4.6cm}}
\hline
\textbf{AI Category} &
\textbf{Purpose} &
\textbf{Subcategories / Model Types} &
\textbf{Representative Applications} \\
\hline

\textbf{1. AI for Interface and Flux Learning} &
Learn fluxes, tendencies, and exchange processes at domain interfaces; enforce conservation and physical consistency &
Physics-guided flux/exchange emulators;  
conservation-preserving interface models;  
hybrid physics--ML surface and boundary-layer schemes;  
learned conservative remapping/harmonization operators &
Air--sea heat and moisture fluxes;  
soil--boundary-layer coupling;  
sea-ice melt/growth exchanges;  
cross-sphere energy and mass balance \\

\textbf{2. AI for Cross-Component Dynamical Modeling} &
Approximate dynamical mappings across components; emulate multi-physics evolution and teleconnected behavior &
Neural operators (FNO, DeepONet, multi-physics operators);  
autoregressive dynamical models;  
learned PDE solvers for interacting spheres;  
continuous operator surrogates for teleconnections &
ENSO dynamics;  
Gulf Stream--atmosphere interaction;  
ocean circulation--carbon feedbacks;  
multi-physics seasonal-to-decadal prediction  \\

\textbf{3. AI for Spatial Interaction Learning \& Geometric Coupling} &
Represent coupling on irregular meshes and multi-sphere geometries; capture spatial dependencies and teleconnections &
Graph neural networks for irregular meshes;  
multi-sphere spatiotemporal graph transformers;  
message-passing teleconnection learners;  
hierarchical and multi-resolution graph operators &
Air--sea interaction patterns;  
land--atmosphere feedbacks;  
wave--ice coupling;  
teleconnections (ENSO, MJO, NAO) \\

\textbf{4. Multi-Domain Foundation Models for Coupling} &
Learn shared latent spaces capturing cross-domain dependencies; support unified multi-sphere prediction &
Cross-domain pretraining;  
multimodal EO \& geophysical encoders;  
latent coupling representation learning;  
FM fine-tuning for unresolved coupling processes &
Atmosphere--land--ocean embeddings;  
climate teleconnections;  
multi-hazard prediction;  
latent representations of coupled variability \\

\textbf{5. Causal \& Explainable AI for Coupling Pathways} &
Identify directional influences, coupling pathways, and regime-dependent feedbacks; interpret ML and physical models &
Lag-aware causal graph discovery (PCMCI/PCMCI+);  
regime-aware causal inference;  
causal representation learning;  
XAI diagnostics for ML couplers &
Soil moisture $\rightarrow$ precipitation feedback;  
SST $\rightarrow$ circulation pathways;  
LAIC;  
magnetosphere--ionosphere--thermosphere (M--I--T) energy propagation \\

\hline
\end{tabular}
\label{tab:coupling_taxonomy_final}
\end{sidewaystable}

\subsection{AI for Interface and Flux Learning}
\label{subsec:ai_flux}

Machine learning is increasingly used to emulate the fluxes and boundary interactions that couple the atmosphere, ocean, land, and cryosphere. A key development in this area is \emph{physics-informed machine learning} (PIML), which incorporates scientific knowledge such as conservation laws, symmetries, and partial differential equations (PDEs) directly into model training \cite{raissi2019physicsinformed,karniadakis2021physics,latrach2024critical}. Rather than fitting data alone, PIML embeds this knowledge into the loss function so that model predictions adhere to the underlying physical constraints. Embedding physics in this manner reduces the need for large training datasets, enhances interpretability, and improves generalization, particularly in regions where observations are sparse or noisy.

In engineering and geosciences, physical knowledge is commonly expressed through ordinary and partial differential equations that govern the evolution of state variables. A generic conservation law can be written as

\begin{equation}
\frac{\partial \phi(x,t)}{\partial t}
+\nabla_{x}\cdot F_{\phi}(x,t)
= S_{\phi}(x,t),
\end{equation}

\noindent
where $\phi$ is a prognostic quantity (e.g., temperature, moisture, momentum), $F_{\phi}$ its flux, and $S_{\phi}$ source, sink, or coupling terms linking different Earth-system components.

Instead of relying purely on data to learn relationships, PINNs leverage the structure of known physics to guide their solutions. A key concept in PINNs is the PDE residual, which measures how well a candidate solution satisfies the governing equations. The residual is defined as

\begin{equation}
\mathcal{R}(x,t)
=
\frac{\partial \phi(x,t)}{\partial t}
+\nabla_{x}\cdot F_{\phi}(x,t)
- S_{\phi}(x,t),
\end{equation}

Ideally, $\mathcal{R}(x,t)=0$, indicating that the approximation satisfies the PDE. During training, the network minimizes the residual across the domain, progressively refining its approximation.

In practice, a physics-informed neural network enforces such governing laws by adding them directly to its training objective. This gives rise to a composite loss function of the form
\begin{equation}
\mathcal{L}
=
\lambda_{\mathrm{data}} \mathcal{L}_{\mathrm{data}}
+\lambda_{\mathrm{Phys}} \mathcal{L}_{\mathrm{Phys}},
\end{equation}
\noindent
where $\mathcal{L}_{\mathrm{data}}$ supervises against observations or high-resolution model output, $\mathcal{L}_{\mathrm{Phys}}$ penalizes the PDE residual defined above.

Beyond the classical PINN formulation, several extensions strengthen physical fidelity in settings relevant to Earth-system coupling. Hamiltonian Neural Networks enforce conservation of energy through their architecture \cite{greydanus2019hamiltonian}, while hidden-physics models aim to recover unknown or partially specified operators directly from data \cite{raissi2019physicsinformed}. Recent reviews, such as the review paper of Wang et al., \cite{wang2025comprehensive}, highlight additional strategies—such as physics-guided regularization, adaptive residual weighting, hybrid numerical–ML formulations, and multi-physics constraint embedding—that improve stability and accuracy in stiff, nonlinear, or multi-scale regimes often encountered at geophysical boundaries. Collectively, these developments demonstrate that explicitly encoding physical structure—whether through conservation-preserving architectures, augmented loss functions, or operator constraints—helps suppress nonphysical drift, maintain balance relationships across components, and enhance the robustness of learned fluxes at key Earth-system interfaces.

\textbf{Implications for Earth system coupling:} In coupled Earth system models, maintaining consistent relationships across components is essential because small local errors can propagate through the coupled system. PIML provides a principled way to constrain AI-based emulators so that atmosphere–ocean, land–atmosphere, and ice–ocean fluxes respect known physics. Chen et al. demonstrated this by embedding the omega equation and water-vapor continuity into a graph neural network for precipitation forecasting, which substantially improved heavy-rainfall predictions by preserving physically consistent links among vertical motion, moisture, and precipitation \cite{Chen2024_GNN_precip}. Similar ideas have been applied to correct interface biases. Wang et al.\ used a CycleGAN to adjust sea-surface temperatures, reducing mean biases and improving the representation of El~Nino variability \cite{Wang2024_SST_GAN}. In addition, hybrid modeling frameworks offer a complementary strategy by pairing physical solvers with learned components. NeuralGCM \cite{Kochkov2024_NeuralGCM} combines a simplified dynamical core with a neural network that learns unresolved fluxes such as cloud–radiation interactions. FaIRGP \cite{bouabid2024fairgp} is a Bayesian machine-learning emulator that combines the physical structure of energy balance climate models with flexible statistical learning to accurately reproduce both global and spatial surface temperature responses under different forcing scenarios. The physically derived core provides a stable backbone, while the learned components capture complex subgrid processes. 

\textbf{Limitations and open challenges:} However, several challenges remain, which highlight the need for stronger physical constraints, systematic stability evaluation, and careful integration of AI components into established Earth system modeling workflows. In particular, learned fluxes may introduce small imbalances that accumulate over long simulations, causing model drift or instability in coupled components \cite{rasp2020weatherbench}. Strongly nonlinear processes near interfaces, including cloud–radiation coupling, sea-ice formation, and mixed-layer turbulence, remain difficult to emulate reliably. Many PIML approaches also struggle with complex or moving boundaries such as the seasonal sea-ice edge or heterogeneous land surfaces. Embedding PDE-based constraints can increase computational cost and may be ineffective when governing equations are only partially known or vary across scales. 

\subsection{AI for Cross-Component Dynamical Modeling}
\label{subsec:ai_multicomponent}

AI-based cross-component dynamical modeling seeks to learn the time evolution of interacting Earth system variables within a unified framework, targeting the joint dynamics of coupled components rather than individual interface processes. This class of approaches encompasses operator-learning methods, continuous-time and generative dynamical models, and hybrid emulators that approximate the evolution of multi-component systems across spatial and temporal scales. In contrast to interface- or flux-focused learning, these methods aim to reproduce the dynamical behavior of full system components, with coupling emerging through the learned state evolution.

Among the most widely adopted approaches are neural operators, which extend conventional neural networks by learning mappings between \emph{function spaces} rather than finite-dimensional vectors \cite{kovachki2023neural}. For dynamical systems governed by partial differential equations, the input is typically a coefficient or state function and the output is the corresponding solution function. Direct discretization of such mappings into fixed grids limits generalization across resolutions and numerical schemes. Neural operators address this limitation by learning resolution-invariant operators defined on bounded domains, enabling evaluation across different discretizations without retraining. Architectures such as the Fourier Neural Operator (FNO) \cite{li2020fourier} implement this idea by learning operator kernels in spectral space, providing efficient approximations of continuous dynamical mappings on heterogeneous spatial domains.

Within this broader landscape of learned dynamical systems, neural operators are particularly attractive for multi-component Earth system settings because they approximate full dynamical operators. As emphasized in \cite{kovachki2023neural}, operator-learning frameworks allow individual components—such as atmospheric flow, ocean circulation, or thermodynamic processes—to be represented by separate learned operators acting on their respective function spaces. These operators can be coupled through the exchange of state variables at each timestep, offering a flexible alternative to traditional operator-splitting and component-coupling strategies. Such modularity has enabled neural operators to replace or augment PDE-based solvers in a range of dynamical and multi-physics applications \cite{li2020fourier,lu2019deeponet}.

\textbf{Implications for Earth system coupling:} Learned dynamical models offer a pathway toward fast and flexible emulation of coupled Earth system components. Neural operators, in particular, provide a common mathematical framework for representing atmosphere, ocean, and land processes, facilitating the exchange of fields across domains and enabling operator-level coupling. Their computational efficiency supports large ensembles and long integrations, expanding the feasibility of uncertainty quantification and probabilistic analysis compared with traditional ESMs. Recent examples illustrate this potential: the Ocean-Linked Atmosphere (Ola) model couples atmospheric and oceanic FNOs to simulate seasonal climate variability and reproduce ENSO-like behavior \cite{Wang2024_SST_GAN}; the Deep Learning Earth System Model (DLESyM) integrates atmospheric, ocean-surface, and radiative neural operators into a compact global surrogate with millennial-scale stability \cite{CresswellClay2025_DLESyM}; and OceanNet demonstrates high-fidelity emulation of regional ocean dynamics at substantial computational speedup \cite{Chattopadhyay2024_OceanNet}. Together, these studies show that operator-based dynamical models can capture both large-scale teleconnections and fine-scale processes when trained on diverse, multiresolution datasets.

Beyond neural operators, a growing body of work explores alternative formulations for learning coupled dynamics. Continuous-time and generative approaches, such as spatiotemporal pyramid flow matching \cite{irvin2025spatiotemporal}, model climate evolution as a stochastic flow in a hierarchical latent space, capturing multiscale spatiotemporal structure without relying on explicit operator formulations. In parallel, physics-aware coupled emulators such as SamudrACE \cite{duncan2025samudrace} directly approximate the joint evolution of three-dimensional atmosphere and ocean states using data generated by physically based climate models. These approaches preserve the structural decomposition and coupling topology of traditional models but do not explicitly enforce governing equations or conservation laws; instead, physical consistency emerges implicitly from the training data and model architecture. Collectively, such methods highlight a spectrum of dynamical modeling strategies, ranging from operator-based surrogates to generative and component-level emulators, all targeting the efficient representation of coupled Earth system dynamics.

\textbf{Limitations and open challenges:} Despite their promise, learned cross-component dynamical models—including neural operators, generative latent-dynamics models, and physics-aware coupled emulators—face several important challenges. When deployed in autoregressive mode, learned dynamical models can accumulate structural errors over long rollouts, leading to drift in energy balance, equilibrium states, or variability patterns \cite{de2023machine}. Architecture-specific biases further affect performance: for example, spectral representations used in operator-based models may suppress small-scale or high-frequency variability, weakening mesoscale interactions and degrading the representation of processes near component interfaces. More broadly, most data-driven dynamical emulators do not explicitly enforce physical conservation laws, leaving them vulnerable to unphysical behavior outside the training regime, particularly during long integrations. Model performance also depends strongly on the coverage, diversity, and stationarity of the training data, raising concerns about robustness under non-stationary forcing, regime shifts, or future climate conditions. Addressing these limitations has motivated the development of physics-regularized operator architectures \cite{beucler2021enforcing}, stability-aware autoregressive training strategies \cite{pathak2022fourcastnet}, and hybrid AI--numerical frameworks that combine learned dynamical components with traditional solvers \cite{raissi2020hidden,li2020fourier}.

\subsection{AI for Spatial Interaction Learning}
\label{subsec:ai_gnn}

Graph Neural Networks (GNNs) are a class of machine learning models specifically designed to operate on graph-structured data, where relationships between entities are as important as the entities themselves~\cite{scarselli2008graph,kipf2017semi}. Until recently, deep learning has been most successful on data with an underlying Euclidean or grid-like structure, such as images or regular latitude--longitude grids~\cite{bronstein2017geometric}. However, many real-world systems, including Earth system, are inherently non-Euclidean and irregular. This has motivated the rapid rise of graph learning as a way to generalize deep learning to non-Euclidean domains~\cite{zhou2020graph}. GNNs provide a principled framework to model such graph data and to exploit the essential relational structure among nodes.

In a GNN, data is represented as nodes (entities) and edges (interactions), allowing for the modeling of complex, irregular domains that go beyond the Euclidean assumptions of standard neural networks. This framework is particularly well-suited for representing dynamical systems, where the state of each node evolves based on its interactions with neighbors. GNNs accomplish this through \emph{message passing}, a process that iteratively updates node representations by exchanging information with adjacent nodes across the edges. This mechanism enables GNNs to learn both local and global dependencies within a system, effectively simulating how influence propagates across a network.

\textbf{Implications for Earth system coupling:} Recent advances demonstrate the potential of GNNs to learn physically meaningful cross-domain interactions directly from data. GraphCast, for example, reformulates the global atmospheric grid as a graph and employs structured message passing to achieve state-of-the-art medium-range forecasts, outperforming leading numerical weather prediction systems \cite{Lam2023_GraphCast}. Building on this idea, GraphDOP extends graph-based prediction to a fully multisphere configuration by constructing a unified graph that integrates atmosphere, ocean, land, and cryosphere observations \cite{Alexe2024_GraphDOP}. Its learned graph dynamics evolve a shared latent state forward in time and reproduce coherent coupled processes such as sea-ice growth affecting boundary-layer stability, SST anomalies modulating large-scale circulation, and soil-moisture variations influencing convective development \cite{Boucher2025_GraphDOP_coupling}.

These developments illustrate how a graph-based representation can provide a unified and flexible mechanism for encoding the heterogeneous geometry of the Earth system. By representing physical relationships explicitly as edges, GNNs naturally capture interactions such as air–sea exchanges, land–atmosphere feedbacks, and ice–ocean coupling without requiring manually specified interface parameterizations. Message passing facilitates the propagation of information across domains and across scales, enabling the model to infer teleconnections and feedback loops directly from observational and reanalysis datasets. Furthermore, the graph formulation integrates seamlessly with multisensor Earth observations—allowing atmospheric, oceanic, land-surface, and cryospheric measurements to be combined within a single relational structure. As a result, GNNs offer a promising pathway toward learned, data-driven coupling mechanisms that can complement and potentially enhance traditional physics-based couplers in Earth system models.

\textbf{Limitations and open challenges:} Despite their promise, several challenges constrain the reliability of GNN-based coupling. Deep message-passing architectures commonly suffer from \emph{oversmoothing}, where repeated aggregation causes node states to become indistinguishable, eroding fine-scale gradients essential for coupled processes \cite{li2018deeper, oono2020graph}. Conservation laws—such as mass, energy, and moisture budgets—are not inherently respected in graph space, and non-conservative learned operators are known to induce long-term drift or unphysical equilibria in geophysical models \cite{thangamuthu2022unravelling}. Constructing an appropriate graph topology remains nontrivial: atmospheric grids, ocean meshes, and land-surface tiles differ in resolution and connectivity, and misaligned or poorly designed graphs can introduce structural biases that propagate through learned dynamics \cite{zhao2024artificial}. Representing teleconnections as long-range edges risks introducing spurious dependencies if such edges are not physically justified. Furthermore, while GNNs perform well in short- to medium-range prediction, their long-horizon stability, drift characteristics, and ability to maintain balanced cross-domain interactions remain largely unexplored; similar issues have been documented for other AI-based dynamical models \cite{de2023machine}. Addressing these challenges will require integrating physical constraints into message passing, developing hybrid graph–physics architectures, and designing evaluation protocols specifically tailored for multi-sphere coupled dynamics.

\subsection{Multi-Domain Foundation Models for Coupling}
\label{subsec:ai_fm}

Foundation models (FMs) have recently emerged as a dominant paradigm in Earth system AI. These models are large-scale neural architectures—often transformer-based—trained on massive and heterogeneous datasets to learn general-purpose representations across variables, domains, and tasks. Unlike traditional machine-learning approaches that require task-specific architectures, foundation models aim to construct a unified latent space in which diverse geophysical processes can be jointly encoded. Their success in natural-language processing and computer vision has motivated analogous efforts in Earth sciences, where abundant reanalysis, satellite, and simulation data provide fertile ground for building high-capacity, multi-domain models.

Early Earth system foundation models, however, focused largely on \emph{single-system} tasks without explicit coupling. Earth-observation FMs such as Prithvi \cite{Prithvi2024}, TerraFM \cite{TerraFM2025}, and TerraMind \cite{TerraMind2025} integrate multimodal satellite imagery for land-cover mapping, change detection, and environmental monitoring. Atmospheric foundation models such as ClimaX \cite{Nguyen2023_ClimaX} learn variable-agnostic representations of reanalysis fields and have demonstrated strong skill in global weather prediction. Yet these single-system models do not represent cross-domain interactions such as atmosphere–ocean fluxes, land–atmosphere feedbacks, or sea-ice–ocean coupling.

\textbf{Implications for Earth system coupling:} Recent advances in foundation models demonstrate a clear shift toward architectures capable of capturing cross-domain interactions directly from data (see table \ref{tab:foundation-models-comparison}). Multi-domain systems such as Aurora \cite{Bodnar2025_Aurora}, trained on more than one million hours of atmospheric, land, ocean-wave, and air-quality data, use shared multivariate embeddings and cross-domain attention mechanisms to internalize relationships resembling classical coupling pathways. AIFS \cite{lang2024aifs} extends this principle by integrating atmosphere, land, and ocean-wave prediction within a unified architecture, exhibiting emergent wave--sea-ice behavior despite lacking an explicit ice component—evidence that coupling-like dynamics can arise from joint training. ORBIT \cite{Wang2024_ORBIT}, scaling to tens of billions of parameters, targets long-range temporal dependencies and teleconnections, learning multi-scale climate variability patterns typically associated with atmosphere–ocean feedbacks and large-scale circulation modes. CAM-NET \cite{hu2025cam} provides a framework that couples the lower atmosphere with the thermosphere–ionosphere, addressing a key limitation of most existing atmospheric foundation models, which primarily focus on the lower atmosphere. Thus, CAM-NET represents an early step toward fast and physically consistent whole-atmosphere Earth system coupling across traditionally separate domains.

A complementary direction is emerging through the fine-tuning of pre-trained foundation models for specialized downstream physical processes. Recent work by \cite{gupta2025finetunin} demonstrates that a 2.3-billion-parameter FM (NASA–IBM’s Prithvi WxC), when fine-tuned on high-resolution atmospheric gravity-wave fluxes, can produce a physically consistent subgrid-scale parameterization for a coarse-resolution climate model. Although gravity waves were not part of the FM’s pre-training objective, the fine-tuned model successfully leverages its learned latent representation of atmospheric evolution to capture vertical momentum transport—an inherently coupling-related process linking tropospheric convection to stratospheric circulation. This result suggests that fine-tuning foundation models can yield downstream components that implicitly encode coupling mechanisms or provide new insights into cross-domain interactions, even for processes absent during foundational training.

Collectively, these developments illustrate that foundation models can act as data-driven surrogate coupled systems: by embedding heterogeneous variables in a shared latent space, exposing them to multi-domain training corpora, and enabling targeted fine-tuning for unresolved processes, they can implicitly encode bidirectional interactions, without requiring manually specified interface parameterizations. This emerging capability positions foundation models as powerful tools for integrated prediction, cross-domain data assimilation, and scientific discovery in coupled Earth system modeling.

\textbf{Limitations and open challenges:} Despite their promise, coupling-oriented foundation models face several unresolved challenges related to representation, generalization, and physical fidelity. Tokenization strategies for multi-domain data often combine incompatible spatial discretizations—such as structured atmospheric grids, unstructured ocean meshes, heterogeneous land tiles, and sparse in-situ observations—risking the loss of physical adjacency and introducing aliasing effects in attention layers. Spatial harmonization across domains with differing resolutions therefore remains an active research problem and directly influences the fidelity of learned cross-domain interactions \cite{zhao2024artificial}. Foundation models also inherit statistical biases from their training datasets, including the under-representation of extremes and rare coupling regimes \cite{camps2025artificial}, which can distort teleconnections and degrade performance for high-impact coupled events. Moreover, most foundation models are trained and evaluated primarily at global scales and for short- to medium-range prediction tasks. As a result, their ability to generalize across spatial scales—from coarse global fields to regional/local observed environments—and across temporal scales—from weather and subseasonal variability to long-term climate dynamics—remains largely untested. Finally, although models such as ORBIT and Aurora demonstrate impressive cross-domain capabilities, systematic evaluation of their long-term stability, conservation properties, and representation of slow feedbacks and multi-decadal coupled processes is still limited. Addressing these challenges will require advances in physics-informed tokenization, conservation-aware architectures, multi-resolution attention mechanisms, hybrid training strategies spanning weather and climate timescales, and benchmark datasets explicitly designed to assess coupled Earth system dynamics.

\subsection{Causal \& Explainable AI for Coupling Pathways}
\label{subsec:ai_causal}

Causal discovery \cite{nogueira2022methods} aims to automatically uncover cause–effect relationships from data, offering a potentially transformative capability for understanding and improving Earth system coupling. Despite its promise, causal discovery remains challenging in practice: many existing approaches rely on strong and often unrealistic assumptions (e.g., independent and identically distributed observations, no unobserved confounding) and are commonly evaluated on overly simplistic synthetic benchmarks, limiting their real-world applicability \cite{brouillard2024landscape}. Classical causal discovery methods \cite{pearl2010causal, peters2014causal, scholkopf2022causality, zanga2022survey} operate on static observational data and generally assume i.i.d. samples. In Earth system sciences, however, this assumption is fundamentally violated. Observations are spatially and temporally correlated, operationally measured at heterogeneous resolutions, influenced by interacting physical processes, and affected by latent or poorly observed variables. As discussed by Brouillard et al. (2024) \cite{brouillard2024landscape}, applying standard causal discovery methods directly to such data can easily produce misleading or spurious conclusions \cite{scholkopf2022causality}.

To overcome these limitations, recent research emphasizes causal inference tailored specifically to time-dependent geophysical systems. Runge (2019) introduced a comprehensive framework for causal discovery in dynamical systems that explicitly accounts for temporal dependencies and distinguishes direct from indirect interactions using conditional independence tests applied across time lags \cite{runge2019inferring}. Building on this foundation, Runge et al. (2023) provided an extensive review of causal inference for time series in Earth and environmental sciences, highlighting methods for handling nonlinearities, high dimensionality, non-stationarity, and observational noise \cite{runge2023causal}. These methods—including PCMCI, PCMCI\texttt{+}, and related constraint-based algorithms—form a methodological backbone for causal discovery in large-scale coupled systems such as the atmosphere–ocean or land–atmosphere domains. Applications of these frameworks have revealed teleconnections, identified early-warning signals of tipping elements, and disentangled direct physical influences from indirect correlations.

Beyond single-dataset approaches, recent work introduces causal discovery methods that integrate information across multiple datasets while accounting for latent contextual variables. Günther et al. (2023) proposed a technique for causal discovery in time series drawn from heterogeneous contexts (e.g., different climate regimes, seasons, or reanalysis versions) by jointly modeling shared causal structure and time-varying latent drivers \cite{gunther2023causal}. Such developments are particularly relevant for Earth system coupling, where processes interact differently under distinct background states (e.g., ENSO versus non-ENSO years, different phases of the Madden–Julian Oscillation).

Structured causal representation learning provides another promising direction. By integrating heterogeneous, multimodal, and spatiotemporal measurements into a latent space shaped by causal constraints, these methods attempt to infer interpretable latent factors aligned with geophysical processes \cite{scholkopf2022causality}. This perspective is valuable for coupled modeling, where many relevant variables (e.g., moisture convergence, dynamical regimes, oceanic subsurface states) are only partially captured by observations or model outputs.

\textbf{Implications for Earth system coupling:} Causal learning provides tools to disentangle the pathways through which anomalies propagate across Earth system components. Temporally explicit approaches such as PCMCI and PCMCI\texttt{+} \cite{runge2019inferring, runge2023causal} enable identification of direct cross-domain interactions—for example, quantifying how SST anomalies influence atmospheric circulation or how soil moisture affects convective activity—beyond what can be inferred from correlations alone. Multi-context causal discovery \cite{gunther2023causal} clarifies how coupling strengths vary with background climate states, supporting analysis of regime-dependent teleconnections or feedback amplification. Causal representation learning \cite{scholkopf2022causality} further enables interpretable latent variables that reflect underlying physical drivers, providing insight into ML-based Earth system models and improving process-level understanding. Causal learning is also beginning to inform explainability in machine-learning-based climate models. Tesch et al. (2023)~\cite{Tesch2023_CausalDL} demonstrated an approach that combines deep learning with causal analysis to investigate land–atmosphere interactions. By training a neural network to predict precipitation from ERA5 fields and then interrogating it with causal attribution techniques, they identified the causal influence of soil moisture on precipitation across regions and seasons, revealing nuanced coupling patterns that complement theory and traditional modeling studies \cite{Tesch2023_CausalDL}. Such approaches highlight how causal learning can help interpret machine-learning surrogates, quantify physical dependencies, and potentially reveal new interactions relevant for improving coupled Earth system models.

\textbf{Limitations and open challenges:} Many causal inference methods rely on assumptions—such as the absence of unobserved confounding or stationarity—that are often violated in climate data \cite{brouillard2024landscape}. In practice, Earth system variables are influenced by numerous latent or partially observed processes (e.g., large-scale circulation modes, subsurface ocean dynamics, anthropogenic forcings), which act as spatiotemporally structured confounders that bias estimated causal relationships. High dimensionality and spatial heterogeneity further complicate conditional independence testing, while observational sparsity in oceanic and cryospheric regions can obscure or distort causal signals. Non-stationary climate regimes introduce context-dependent causal structures that remain difficult to disentangle, even for state-of-the-art methods. Causal representation learning inherits the limitations of the underlying predictive model, and causal attribution applied to deep networks may misrepresent physical influences when the neural model itself is biased \cite{Tesch2023_CausalDL}. Addressing these challenges will require hybrid causal–physical approaches, improved treatment of latent confounders, rigorous uncertainty quantification, physically informed constraints, and benchmark datasets specifically designed for causal discovery in coupled geophysical systems.

%%%%%%%%%%%%%%%%%%%%%%%%%%%%%%%%%%%%%%%%%%%%%%%%%
% 6. Evaluation Section
%%%%%%%%%%%%%%%%%%%%%%%%%%%%%%%%%%%%%%%%%%%%%%%%%

\section{Evaluation of AI Couplers}
\label{sec:evaluation}
Ensuring the reliability of AI-based couplers requires evaluation protocols that go beyond standard single-domain ML verification. Because coupled Earth system components exchange fluxes, tendencies, and boundary conditions, even small inconsistencies in one domain can propagate across the interface and accumulate into large-scale biases or instabilities over long simulations. Physical consistency must therefore serve as a primary validation criterion. Recent studies have shown that non-conservative or weakly constrained ML components can produce artificial energy sources and sinks, moisture imbalance, or spurious drift when embedded in climate models \cite{thangamuthu2022unravelling}. Conservation diagnostics should evaluate energy, mass, momentum, and moisture budgets jointly across components, including interface flux matching, closed global integrals, and balance constraints such as hydrostatic and geostrophic adjustment.

Long unforced or weakly forced integrations remain indispensable for detecting emergent pathologies in AI-driven coupled dynamics. Such experiments reveal whether neural operators, graph-based models, or cross-domain foundation models accumulate drift in sea-surface temperature (SST), top-of-atmosphere radiation, large-scale circulation, ENSO variability, or sea-ice extent. Ensemble-based experiments and perturbed-physics tests can probe sensitivity to stochasticity, internal variability, and structural perturbations, while targeted stress tests---e.g., strong atmosphere–ocean feedback episodes, blocking regimes, or rapid cryosphere transitions---help expose failure modes that may not be visible in short hindcasts.

Skill assessment for AI couplers must also consider multivariate and cross-domain correlations, rather than evaluating each component in isolation. Metrics should capture atmosphere–ocean covariance fidelity, land–atmosphere flux coherence, and the accuracy of teleconnection structures that link distant regions of the Earth system. Recent work emphasizes diagnostics for interfacial dynamics, including flux-transmittance indices, operator consistency tests, energy-transfer spectra, and response-function analyses that quantify whether AI components preserve the correct pathways and magnitudes of cross-domain interaction \cite{russell2018metrics}. Evaluating whether AI-generated tendencies produce physically meaningful emergent patterns—such as ENSO teleconnections or monsoon–land feedbacks—is increasingly important as couplers become more complex and domain-general.

A persistent challenge is the scarcity of benchmark datasets explicitly designed for evaluating coupled AI systems. Resources such as ClimSim \cite{yu2023climsim}, FourCastNet \cite{pathak2022fourcastnet}, WeatherBench2 \cite{potter2023weatherbench2}, and CMIP-based archives provide valuable training and validation data, but they predominantly emphasize atmospheric fields. Progress requires multi-sphere benchmark suites with aligned ocean, land, cryosphere, and biogeochemical states, including consistent fluxes and high-frequency boundary tendencies suitable for testing cross-domain fidelity.

Uncertainty quantification (UQ) is particularly critical for AI couplers because uncertainties propagate through interacting components in nonlinear ways. Bayesian neural operators, ensemble-based neural surrogates, and probabilistic transformer architectures are emerging approaches for representing structured uncertainty in spatiotemporal fields \cite{li2023bayesiannos, kovachki2023fno_review}. Data-assimilation frameworks enriched with machine learning---including hybrid ensemble–variational methods and ML-calibrated error covariance models---enable systematic propagation of uncertainty between the coupled domains they integrate \cite{bocquet2020uncertainty, Cheng2023_ML_DA_UQ_review}. A central challenge is ensuring that uncertainty estimates remain physically coherent: uncertainties in one domain should induce consistent uncertainty propagation across shared interfaces, such as turbulent heat fluxes, freshwater exchange, or momentum transfer.

Ultimately, evaluation of AI couplers must be treated as a systems-level problem. Robust assessment requires verifying physical consistency, long-term stability, multivariate relationships, interface fidelity, and uncertainty propagation simultaneously. Developing community benchmarks, standardized diagnostics, and shared experimental protocols will be essential to establish trust in next-generation AI-driven Earth system coupling frameworks.

%%%%%%%%%%%%%%%%%%%%%%%%%%%%%%%%%%%%%%%%%%%%%%%%%
% 7. Conclusions
%%%%%%%%%%%%%%%%%%%%%%%%%%%%%%%%%%%%%%%%%%%%%%%%%

\section{Conclusions}
This review has examined how artificial intelligence is reshaping the representation, interpretation, and prediction of coupled processes in the Earth system. AI methods now span interface emulators, neural operators, graph-based interaction models, multi-domain foundation models, and causal discovery frameworks, each offering new ways to capture the pathways through which energy, mass, and momentum flow across spheres. These approaches complement traditional Earth System Models by enabling flexible parameterizations, accelerating multicomponent simulations, and exploiting heterogeneous observations that extend beyond the atmosphere–ocean–land–cryosphere system.

A central theme of this review is that coupling constitutes both the physical structure of the Earth system and the organizing principle for next-generation AI models. AI has demonstrated clear potential to enhance cross-domain flux estimates, emulate multi-component dynamics, identify causal interactions, and support prediction in emerging coupling regime. At the same time, substantial challenges remain. Ensuring physical consistency, long-term stability, and strict conservation in AI-driven components is essential, as is developing rigorous evaluation protocols that reflect the multivariate and multiscale nature of coupled dynamics. Robust uncertainty quantification and interpretability remain key obstacles for deploying AI systems in scientific or operational settings.

Looking forward, the convergence of AI, high-performance computing, and traditional numerical modeling within digital-twin frameworks offers a pathway toward integrated systems in which data assimilation, simulation, and prediction are tightly linked. Realizing this potential will require close collaboration across disciplines and careful design of hybrid architectures that combine physical fidelity with data-driven flexibility. The long-term objective is to develop trustworthy, physically grounded AI-enabled Earth system models that not only reproduce complex couplings but also advance our understanding of the mechanisms driving variability and change in the Earth system.

\section*{Data availability}
Data sharing not applicable to this article as no datasets were generated or
analysed during the current study.

\section*{Author contributions}

Dr. Maria Kaselimi conceived the study and contributed to the machine-learning–based methodology, model analysis, and synthesis of AI approaches. 

\noindent
Dr. Anna Belehaki contributed to the Earth system coupling perspective, domain interpretation, and physical context, particularly regarding atmosphere–ionosphere–thermosphere interactions. 

\noindent
Both authors contributed to the literature review, manuscript writing, and revision, and approved the final version of the manuscript.

\noindent
The authors declare no competing interests.

%%%%%%%%%%%%%%%%%%%%%%%%%%%%%%%%%%%%%%%%%%%%%%%%%
% Bibliography
%%%%%%%%%%%%%%%%%%%%%%%%%%%%%%%%%%%%%%%%%%%%%%%%%

{\small
\bibliography{sn-bibliography}
} 

\end{document}